# Cultural Convergence
# Insights into the behavior of misinformation networks on Twitter

**Liz McQuillan, Erin McAweeney, Alicia Bargar, Alex Ruch**
**Graphika, Inc.**
**{firstname.lastname@Graphika.com}**

## Abstract

How can the birth and evolution of ideas and communities in a network be studied over time? We use a multimodal pipeline, consisting of network mapping, topic modeling, bridging centrality, and divergence to analyze Twitter data surrounding the COVID-19 pandemic. We use network mapping to detect accounts creating content surrounding COVID-19, then Latent Dirichlet Allocation to extract topics, and bridging centrality to identify topical and non-topical bridges, before examining the distribution of each topic and bridge over time and applying Jensen-Shannon divergence of topic distributions to show communities that are converging in their topical narratives.

## 1 Introduction

The COVID-19 pandemic fostered an information ecosystem rife with health mis- and disinformation. Much as COVID-19 spreads through human interaction networks (Davis et al., 2020), disinformation about the virus travels along human communication networks. Communities prone to conspiratorial thinking (Oliver and Wood, 2014; Lewandowsky et al., 2013), from QAnon to the vaccine hesitant, actively spread problematic content to fit their ideologies (Smith et al., 2020). When social influence is exerted to distort information landscapes (Centola & Macy, 2007; Barash, Cameron et al., 2012; Zubiaga et al., 2016; Lazer et al., 2018, Macy et al., 2019), large groups of people can come to ignore or even counteract public health policy and the advice of experts. Rumors and mistruths can cause real world harm (Chappell, 2020; Mckee & Middleton, 2019; Crocco et al., 2002) – from tearing down 5G infrastructure based on a conspiracy theory (Chan et al., 2020) to ingesting harmful substances misrepresented as "miracle cures" (O'Laughlin, 2020). Continuous downplaying of the virus by influential people likely contributed to the climbing US mortality and morbidity rate early in the year (Bursztyn et al., 2020).

Disinformation researchers have noted this unprecedented 'infodemic' (Smith et al., 2020) and confluence of narratives. Understanding how actors shape information through networks, particularly during crises like COVID-19, may enable health experts to provide updates and fact-checking resources more effectively. Common approaches to this, like tracking the volume of indicators like hashtags, fail to assess deeper ideologies shared between communities or how those ideologies spread to culturally distant communities. In contrast, our multimodal analysis combines network science and natural language processing to analyze the evolution of semantic, user-generated, content about COVID-19 over time and how this content spreads through Twitter networks prone to mis-/disinformation. Semantic and network data are complementary in models of behavioral prediction, including rare and complex psycho-social behaviors (Ruch, 2020). Studies of coordinated information operations (Francois et al., 2017) emphasize the importance of semantic (messages' effectiveness) and network layers (messengers' effectiveness).

We mapped networks of conversations around COVID-19, revealing political, health, and technological conspiratorial communities that consistently participate in COVID-19 conversation and share ideologically motivated content to frame the pandemic to fit their overarching narratives. To understand how disinformation is spread across these communities, we invoke theory on cultural holes and bridging. Cultural bridges are patterns of meaning, practice, and discourse that drive a propensity for social closure and the formation of clusters of like-minded individuals who share few relationships with members of other clusters except through bridging users who are omnivorous in their social interests and span multiple social roles (Pachucki & Breiger, 2010; Vilhena et al 2014). While cultural holes/bridges extend our ability to comprehend how mis-/disinformation flow in networks and affect group dynamics, this work has mostly focused on pro-social behavior. It has remained an open question if cultural bridges operate similarly with nefarious content as with beneficial information, as the costs and punishments of sharing false and potentially harmful content differ (Tsvetkova and Macy, 2014; 2015). This work seeks to answer that question by applying these theories to a network with known conspiratorial content.

We turn to longitudinal topic analysis to track narratives shared by these groups. While some topical trends around the COVID-19 infodemic are obvious, others are nuanced. Detecting and tracking subtle topical shifts and influencing factors is important (Danescu-Niculescu-Mizil et al., 2013), as new language trends can indicate a group is adopting ideas of anti-science groups (e.g., climate deniers, the vaccine hesitant). Differences between community interests, breadth of focus, popularity among topics, and evolution of conspiracy theories can help us understand how public opinion forms around events. These answers can quantify anecdotal evidence about converging conspiratorial groups online.

To capture these shifts, we analyze evolving language trends across different communities over the first five months of the pandemic. In (Bail, 2016), advocacy organizations' effectiveness in spreading information was analyzed in the context of how well their message 'bridged' different social media audiences. A message that successfully combined themes that typically resonated with separate audiences received substantially more engagements. We hypothesize topical convergence among communities discussing COVID-19 will 1) correlate with new groups' emergence and 2) be precursed by messages combining previously disparate topics that receive high engagement.

## 2 Methodology

### 2.1 Data

We use activity[1] from 57,366 Twitter accounts, recorded January-May 2020 and acquired via the Twitter API. Non-English Tweets[2] were excluded from analysis. We applied standard preprocessing, removing punctuation and stop words[3], as well as lemmatization of text data. In total, 81,348,475 tweets are included.

### 2.2 Overview of Analysis

We use Latent Dirichlet Allocation (LDA) combined with divergence analysis (Jensen-Shannon; Wong & You, 1995) to identify topical convergence as an indication of the breadth and depth of content spread over a network. We expect that as communities enter the conversation, topical distributions will reflect their presence earlier than using network mapping alone. We first mapped a network, using COVID-19 related seeds from January-May 2020. We then applied LDA to the tweet text of all accounts that met inclusion criteria (discussed below), tracked the distribution of topics both overall and for specific communities, calculated bridging centrality measures to identify topics that connect otherwise disconnected groups.

### 2.3 Network Mapping

[1] Account activity includes tweet text, timestamps, follows and engagement based relationships over accounts.
[2] Language classification was done with pycld3 (https://github.com/bsolomon1124/pycld3)

[3] Punctuation and stopword removal was done with spaCy (https://github.com/explosion/spaCy/blob/master/spacy/lang/punctuation.py; https://github.com/explosion/spaCy/blob/master/spacy/lang/en/stop_words.py)

We constructed five network maps that catalogue a collection of key social media accounts around a particular topic – in this case, COVID-19. Our maps represent cyber-social landscapes (Etling et al., 2012) of key communities sampled from representative seeds (Hindman & Barash, 2018) of the topic to be analyzed. We collect all tweets containing one or more seed hashtags and remove inactive accounts (based on an activity threshold described below). We collect a network of semi-stable relationships (Twitter follows) between accounts, removing poorly connected nodes using k-core decomposition (Dorogovtsev et al., 2006). We apply a technique called "attentive clustering" that applies hierarchical agglomerative clustering to assign nodes to communities based on shared following patterns. We label each cluster using machine learning and human expert verification and organize clusters into expert identified and labeled groups.

## 2.4 Map Series Background

Our maps can be seen as monthly "snapshots" of mainstream global conversations around coronavirus on Twitter. These maps were seeded on the same set of hashtags (#CoronavirusOutbreak, #covid19, #coronavirus, #covid, "COVID19", "COVID-19"), to allow a comparison in network structure and activity over time. For the first three months, accounts that used seed hashtags three or more times were included; for subsequent months, COVID-19 conversations became so ubiquitous that we benefited from collecting all accounts that used seed hashtags at least once. Mapping the cyber-social landscape around hashtags of interest, rather than patterns of how content was shared, reveals the semi-stable structural communities of accounts engaged in the conversation.

## 2.5 Topic Modeling

After extracting and cleaning the tweets' text, we create an LDA model. LDA requires a corpus of documents, a matrix of those documents and their respective words, and various parameters and hyperparameters[4]. LDA, by design, does not explicitly model temporal relationships. Therefore, our decision to define a document as weekly collections of tweets for a given user allows for comparison of potentially time-bounded topics.

We use Gensim (Řehůřek & Sojka, 2010) to build and train a model, with the number of topics $K$= 50, and two hyperparameters $\alpha$ = 0.253 and $\beta$ = 0.946[5], all tuned using Optuna[6] to optimize both for coherence score and human interpretability. LDA then maps documents to topics such that each topic is identified by a multinomial distribution over words and each document is denoted by a multinomial distribution over topics. While variants of LDA include an explicit time component, such as Wang and McCallum's (2006) Topics over Time model or Blei and Lafferty's (2006) Dynamic Topic Model, these extensions are often inflexible when the corpus of interest has large changes in beta distributions over time, and further impose constraints on the time periods which would have been unsuitable for this analysis.

Applying LDA with $K$ = 50 yielded topics (see Appendix A) that can be distinguished by subject matter expert analysis. A substantial portion of topics related to COVID-19, politics at national and international levels, international relations, and conspiracy theories. We also identified a small subset of social media marketing and/or peripheral topics, which covered animal rights, inspirational quotes (topic 10), and follow trains (topics 23 and 32).

[4] This research defines a document as *all* tweets for a given account during a single week (Monday-Sunday) concatenated and uses a matrix of TF-IDF scores in place of raw word frequencies.

[5] With $\alpha$ being a measure of document topic density (as $\alpha$ increases so does the likelihood that a document will be associated with multiple topics), and $\beta$ being a measure of topic word density (as $\beta$ increases so does the likelihood that topics use more words from the corpus). LDA is a generative, latent variable model which assumes all (observed) documents are generated by a set of (unobserved) topics, hyperparameters have a noticeable impact on the assignment of documents to topics.

[6] https://optuna.readthedocs.io/en/stable/index.html

### 2.6 Cultural Bridge Analysis

In addition to topics derived from LDA, we extracted hashtags, screen names, and URLs from tweets as possible cultural content that bridges communities. For each week, we construct an undirected multimodal graph where each node represents a cluster, topic, or content. Clusters have edges to topics with a weight representing average topic representativeness among cluster members, and edges to cultural content with a weight according to the number of unique cluster members using said content divided by maximum usage of an artifact of that type. After constructing the graph, we calculate nodes' bridging centrality, respect to neighbors' degrees to better identify nodes spanning otherwise disconnected clusters (Hwang et. al, 2006). We select the top 20 bridges by node type for further investigation.

## 3 Results

### 3.1 Five Months of Maps

All five maps are visibly multipolar, with densely clustered poles and sparsely populated centers[7]. This indicates that strongly intra-connected communities are involved in the conversation despite its global scope and each geographic and ideological cluster has its own insular sources of information - there is a dearth of shared, global sources of information.

This lack of cohesion or "center" to the network is not unusual for a global conversation map, as dense clusters often denote national sub-communities. However, the first coronavirus map covers a time period in which the outbreak was largely contained to China, with the exception of cases reported in Thailand toward the end of January.[8] During this time, the Chinese government and its various media outlets would be expected to form that core informational source. As noted in previous studies and reporting of health crises, consensus from the medical and science community challenging, particularly in online spaces.[9] This in turn creates voids of both information and data for bad actors to capitalize up.

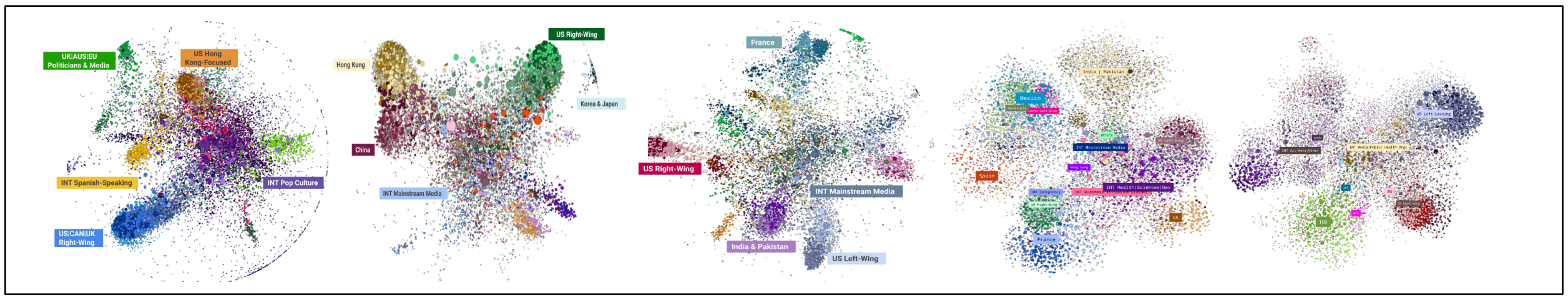

*Figure 1. January-May 2020 Network Maps*

### 3.2 Cultural Convergence Case Study: QAnon

Our pipeline yielded a large set of results detailing the cultural convergence of clusters over time in the Covid-19 network maps and the topical bridges that aided this convergence. The QAnon community is a particularly apt example since researchers studying this topic have qualitatively noted the recent acceleration of QAnon membership since the beginning of the pandemic and its convergence with other online communities. The aim of the QAnon case study is to illustrate the power of this cultural convergence method and reflect similar results we see in numerous groups across our collection of Covid-19 map clusters.

[7] We visualize the map network using a force-directed layout algorithm similar to Fruchterman-Rheingold (1991) – individual accounts in the map are represented by spheres, pushed apart by centrifugal force and pulled together by spring force based on their social proximity assign a color to each account based on its parent cluster.

[8] https://web.archive.org/web/20200114084712/https://www.cdc.gov/coronavirus/novel-coronavirus-2019.html

[9] https://www.usatoday.com/story/news/health/2019/04/23/vaccine-measles-big-pharma-distrust-conspiracy/3473144002/

### 3.3 QAnon and Covid-19

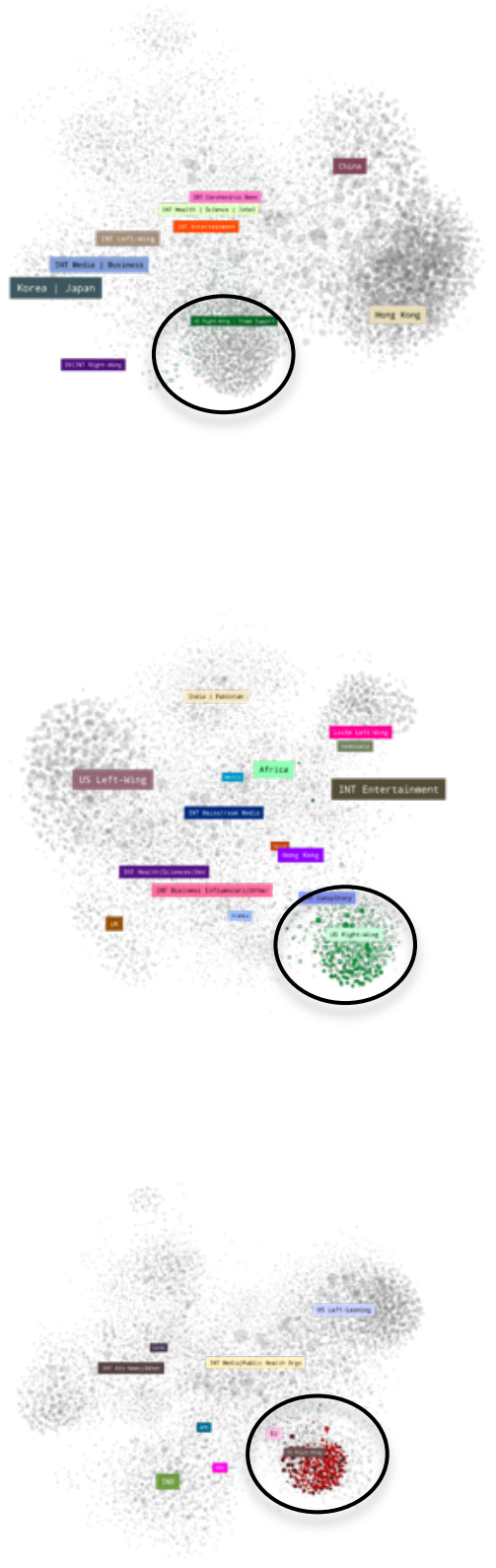

*Fig.2 QAnon in February, April, and May.*

QAnon is a political conspiracy theory formed in 2017 (LaFrance, 2020). The community maintains that there is a secret cabal of elite billionaires and democratic politicians that rule the world, also known as the "deep state." As the theory goes, Donald Trump is secretly working to dismantle this powerful group of people. Given Trump's central role and the vilification of democratic politicians, this far-right theory is most commonly adopted by Trump supporters.

Researchers that study the QAnon community have hypothesized an accelerated QAnon membership since the beginning of the pandemic (Breland & Rangarajan, 2020). Conspiratorial threads, like the "plandemic" that asserts COVID-19 was created by Bill Gates and other elites for population control, are closely tied to QAnon's worldview. Our results support the hypothesis that over the course of the pandemic, the QAnon community has not only grown in size, but the QAnon content and concepts have taken hold in other communities. Further, our results show that this fringe ideology has spread to broader, more mainstream groups.

The QAnon community started as a fringe group that has notably gained momentum both in our COVID-19 networks and in real life. The first "Q" congressional candidate was elected in Georgia and is favored to win the House seat (Itkowitz, 2020). QAnon has attracted mainstream coverage as well (Strauss, 2020; CBS News, 2020; LaFrance, 2020). In accordance with these real-world impacts, our time series of maps illustrate a similar process of QAnon spreading to the mainstream online, both in network and in narrative space. Prior to the pandemic, QAnon was a fringe conspiracy concentrated on the network periphery of Trump support groups. As Figure 3 illustrates, the QAnon community was less than 3% of the conversation around COVID-19 in February 2020. By April 2020, this community comprised almost 5% of the network, and was composed of increasingly dense clusters (in February the QAnon group had a heterophily score of .09 whereas in May scores for QAnon clusters ranged from 5.57-20.77). At the same time, there is consistent growth in the number of documents assigned to the QAnon topic (28) from 12/2019 to 5/2020:

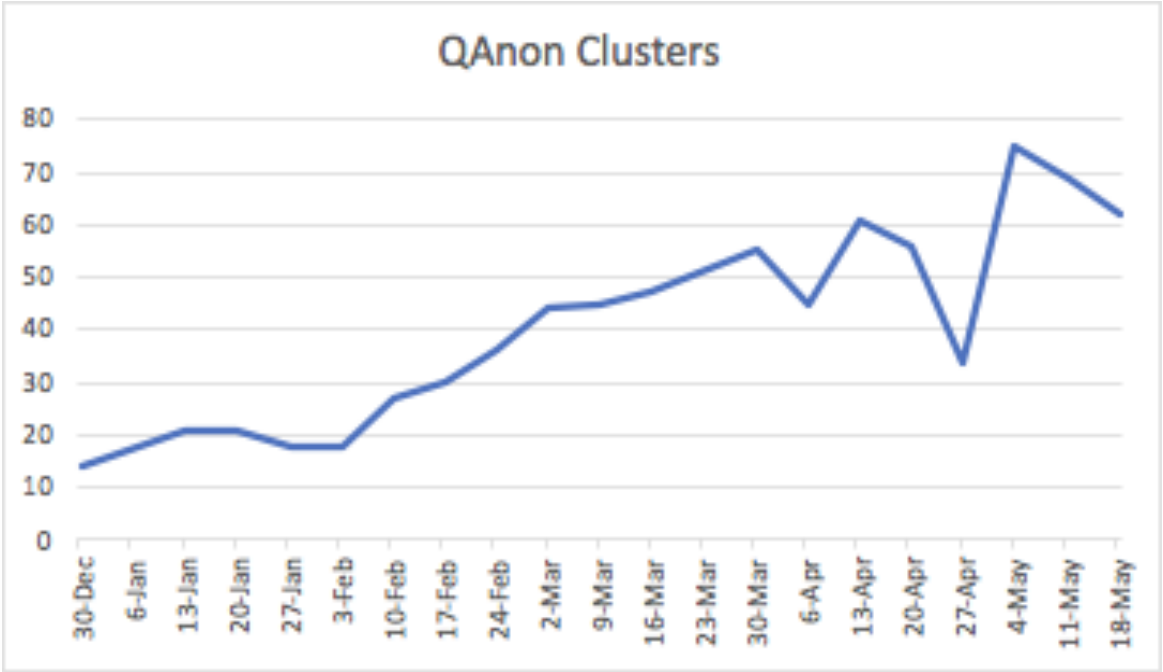

*Fig 3. Growth of QAnon Clusters Over Time*

Our cultural convergence method depicted the topics that supported this shift and the groups that adopted QAnon concepts and content over time. That process within a few groups is outlined below.

### 3.4 QAnon and Right-Wing Groups

All of the network maps in Figure 2 feature a large US Right-Wing group, indicating that this group was prominently involved in the coronavirus conversation. Both these maps demonstrate what we refer to as mega clusters, loud online communities with a high rate of interconnection. This interpretation is supported by both topical and non-topical bridging centrality measures. For example, after the assassination of Iranian major General Qasem Soleimani in early January, we see US/Iran Relations (Topic 28) narratives acting as a bridge between several US Right-Wing communities (US|CAN|UK Right-Wing in blue above) who use this topic to connect with each other as well as with accounts who, in later months, we will see come together to form distinct QAnon and conspiracy theory communities. While documents from these groups make up a relatively small proportion of all documents within the topic at first, they become more prevalent over time: the US/Iran relations topic (Topic 28) becomes the dominant bridge in numerous and geographically diverse clusters by mid-March and continues to hold that spot through mid-April.

We also find that US Right-Wing clusters converge topically at the same time as US/Iran becomes more of a bridge: at the start of our analysis the similarity between the topical distribution of these clusters is relatively low (for example, the US Trump|MAGA Support I and US Trump Black Support|Pro-Trump Alt-media clusters have a divergence score of 0.3312 during the week of 12/30/2019 indicating minor overlap). By 3/30/2020, the same clusters have a divergence score of 0.5697 and by 5/18/2020 their divergence score reaches 0.6986, over double compared to three months earlier.

Beginning the week of March 16th, we see QAnon making significant gains in narrative control of the network, with QAnon related narratives (i.e. Topic 11) acting as one of the strongest bridges (maximum bridging centrality for Topic 11 is 0.4794) between communities including conspiracy theorists on both the Right and Left, and the US Right Wing. This persists throughout the remainder of March and into mid-late April, with April being the first month we see more than one distinct QAnon community detected in our network maps. Again, this engagement is supported by the convergence of Topic 49 over the analysis period. In January these groups have relatively minor similarities in their topic distributions (0.3835), though these similarities become much larger by mid-March (0.6619) and continue to increase through the end of the analysis period in mid-May (0.7150).

It is notable that before we are able to detect distinct QAnon communities through network mapping, they are already making connections to more established communities within the COVID network through this content area. In terms of overall map volume, the Right-Wing group was the third largest in the January map, with 15.9%, and the largest in the February map, with 22.6%. Overall, since March, Left-leaning and public health clusters gained a foothold in the conversation. Our results show the Right-wing groups were pushed to the periphery of the maps since March and at the same time subsumed by the fringe QAnon conspiracy.

By the end of April, Hong Kong protest related narratives (Topic 8) replaced QAnon as the strongest bridge. However, given that April and May each have several distinct QAnon communities, it is possible that QAnon topical content is being contained within those newly coalesced communities rather than bridging formerly distinct communities while Hong Kong protest related narratives are more culturally agnostic. This is supported by the lack of MAGA/US Right Wing clusters in later months and relational increase in QAnon clusters.

### 3.5 QAnon and the Alt-News Networks

In the beginning of the pandemic, the volume of clickbait “news” sites, which tend to spread unreliable and sensational COVID-19 updates, appeared between January to February (the group made-up of these accounts and their followers increased from none in January to close to 6% of the map in February). During this time many accounts from the alt-media and conspiratorial clusters exclusively and constantly tweeted about coronavirus, some including the topic in their profile identity markers. This is mirrored on other platforms, for instance on Facebook where we saw groups changing their names to rebrand into COVID-19 centric groups, while they were previously groups focused on other political issues. These alternative "news" accounts are preferred news sources for clusters that reject "mainstream news." This sentiment is reflected in both alt-right and alt-left clusters that carry anti-establishment and conspiratorial views. Notably, these two groups also engage with and share Russian state media, like RT.com, regularly. In the February map the dedicated Coronavirus News group (5.8%), which is composed of accounts that follow these alternative and often poorly fact-checked media sources, was almost equivalent in size to the group of those following official health information sources (7.4%).

By March, conspiratorial accounts and alt-right news sources like Zero Hedge and Breitbart were missing from the top mentions across this map and were replaced by influential Democrats such as Bernie Sanders and Alexandria Ocasio-Cortez and left-leaning journalists such as Jake Tapper and Chris Hayes. In line with the trajectory of more mainstream voices becoming engaged in the conversation as the outbreak progressed, fringe voices became less influential in our maps over time. These changes could represent a mainstreaming of coronavirus conversation, which in turn makes the center of the map more highly concentrated, or the reduced share of a consolidated US Right-Wing community discussing coronavirus online.

Our results show that preference for QAnon concepts converge on both ends of the political ideological spectrum. This supports the "horseshoe theory," frequently postulated by political scientists and sociologists, that the extremes of the political spectrum resemble one another rather than being polar opposites on a linear political continuum. In March-May, the QAnon topic was an important bridge between alt-left and alt-right clusters (the lowest divergence being 0.73 on 3/30/20 and 0.50 on 5/18/2020 in the same cluster). The topic remained a dominating bridge each week across our maps, suggesting that this bridge is a strong tie between the alt-left and alt-right groups independent of the pandemic. The alt-left is often vocal on anti-government, social justice, and climate justice topics. From our results we can also see that QAnon is also a topical bridge between accounts following alt-left journalists and environmental and climate science organizations (0 .77 on 3/30/20 and 0.75 on 5/18/2020). This suggests that while the far-right is easily drawn to QAnon content because of the anti-liberal bend, there are other channels, such climate activism, that act as channels for the QAnon conspiracy to spread.

## 4 Conclusion

The multi-modal approach to cultural convergence helps us better understand the highly dynamic nature of overlapping conspiratorial strands. Our findings highlight that conspiratorial groups are not mutually exclusive and this approach models some of the driving forces behind these convergences. The QAnon case study is a fraction of the results yielded by this approach and highlights important insights into cultural and topical interconnections between online groups. Further work will explore the glut of results generated by applying this approach to the ongoing time series mapping of COVID-19. Other topics, such as a time series of maps around the recent Black Lives Matter protests will also be explored.

## A Appendix

### A.1 Topics

| Topic | Expert Assigned Label | Words |
|---|---|---|
| 0 | Right-Wing, Anti-CCP, Steve Bannon | warroompandemic, bannon, jasonmillerindc, warroom2020, raheemkassam, realdonaldtrump, ccp, jackmaxey1, robertspalding, steve, chinese, pandemic, ccpvirus, war, vog2020 |
| 1 | Christian, Hindu | god, saint, ji, ✍️, TRUE, lord, spiritual, holy, jesus, rampal, maharaj, bible, knowledge, kabir, sant |
| 2 | Food, Trump, Covid | tasty, recipe, recipes, food, foodie, cookies, chicken, delicious, realdonaldtrump, homemade, pandemic, ios, chocolate, salad, insiderfood |
| 3 | Tech Industry, ML Interest | ai, data, cybersecurity, read, business, security, iot, tech, technology, bigdata, digital, machinelearning, 5, free, – |
| 4 | Covid, Health, Human Rights | climate, women, global, change, pandemic, children, vaccine, study, youtube, crisis, human, –, read, un, research |
| 5 | Nazi Germany, Holocaust | auschwitzmuseum, born, auschwitz, jewish, february, 1942, polish, incarcerated, deported, march, 1944, jew, 1943, raynman123, breakingnews |
| 6 | US-Iran relations | iran, iranian, soleimani, iraq, regime, heshmatalavi, irans, tehran, war, iraqi, iranians, killing, killed, irgc, realdonaldtrump |
| 7 | Trump Support Accounts, Qanon | tippytopshapeu, mevans5219, dedona51, philadper2014, donnacastel, f5de, jamesmgoss, fait, thepaleorider, iam, ☣️, ecuador, newspaper, realdonaldtrump, unionswe |
| 8 | Hong Kong Protests | hong, kong, chinese, police, ccp, hongkong, wuhan, hk, communist, taiwan, beijing, solomonyue, chinas, party, human |
| 9 | Pro-Russia, Russia news | syria, iran, war, military, turkey, russia, israel, russian, turkish, forces, iraq, syrian, idlib, rtcom, killed |
| 10 | Inspirational Quotes, SMM | love, thinkbigsundaywithmarsha, quote, 4uwell, joytrain, joy, peace, kindness, motivation, mindfulness, things, quotes, success, mentalhealth, happy |
| 11 | Democrat-focused Qanon | □, realdonaldtrump, f1faf1f8, patriots, ❤️, potus, america, democrats, god, biden, follow, ⭐, obama, joe, dems |
| 12 | Trump Support Accounts, Conservative Influencers, Christians | realdonaldtrump, biden, joe, realjameswoods, chinese, democrats, americans, american, america, flu, national, charliekirk11, god, bernie, house |
| 13 | India/Pakistan Covid, Muslim | pakistan, sindh, india, minister, khan, pm, imrankhanpti, govt, indian, kashmir, allah, imran, corona, karachi, pakistani |
| 14 | Christian Qanon, Qanon influencer | realdonaldtrump, q, inevitableet, cjtruth, prayingmedic, qanon, stormisuponus, lisamei62, potus, eyesonq, god, juliansrum, karluskap, thread, |
| 15 | Covid News | deaths, total, reports, county, confirmed, positive, death, update, pandemic, york, reported, police, number, bringing, city |
| 16 | K-pop | thread, unroll, hi, read, find, hello, asked, follow, be, kenya, sorry, dm, thanks, bts, retweet |
| 17 | Democratic Primaries, Candidates | bernie, biden, sanders, berniesanders, joe, vote, warren, joebiden, campaign, democratic, bloomberg, be, candidate, voters, primary |
| 18 | Covid General | pandemic, lockdown, workers, stay, social, positive, amid, food, care, minister, dr, spread, crisis, emergency, testing |
| 19 | Malaysia News | malaysia, nstnation, fmtnews, staronline, minister, pm, malaymail, malaysian, nstonline, mahathir, nstworld, muhyiddin, dr, malaysians, kkmputrajaya |
| 20 | South Africa News | africa, south, lockdown, african, sa, news24, minister, africans, cyrilramaphosa, zimbabwe, black, cape, ramaphosa, anc, police |
| 21 | Missouri Reps, Libertarians, Bitcoin | washtimesoped, tron, realdonaldtrump, truthraiderhq, trx, govparsonmo, chinese, hawleymo, democrats, hong, kong, czbinance, rephartzler, biden, |
| 22 | Politicsl Interest, News Platforms | smartnews, googlenews, yahoonews, franksowa1, realdonaldtrump, trimet, biden, obama, americans, aol, tac, tic, house, white, pandemic |
| 23 | Trump Train | §, 18, code, warnuse, 2384, seditious, , conspiracy, viccervantes3, ☠️, realdonaldtrump, 1962, rico, f9a0, f680vicsspaceflightf680 |
| 24 | Canada General Interest | canada, trudeau, canadians, justintrudeau, canadian, ontario, alberta, pandemic, minister, jkenney, cdnpoli, ford, care, pm, fordnation |
| 25 | Trump Support, Jewelry | fernandoamandi, necklace, sariellaherself, earrings, realdonaldtrump, police, kong, jewelry, hong, catheri77148739, bracelet, handmade, turquoise, america, pendant |
| 26 | Wikileaks, Isreal-Palestine Relations | israel, palestinian, israeli, assange, gaza, palestinians, julian, palestine, swilkinsonbc, occupation, children, prison, rights, war, wikileaks |
| 27 | Alt-Right News | nicaragua, zyrofoxtrot, banneddotvideo, allidoisowen, dewsnewz, iscresearch, realdonaldtrump, f4e2, offlimitsnews, libertytarian, f1faf1f8, ortega, infowars, f53b, |
| 28 | US-Iran relations, US politicians | iran, realdonaldtrump, iranian, soleimani, democrats, impeachment, american, america, pelosi, obama, terrorist, war, killed, house, iraq |
| 29 | India Celebs, Entertainment | ❤️, master, , rameshlaus, tn, chennai, birthday, happy, fans, actorvijay, f60e, tweets, film, tamilnadu, thala |
| 30 | US coronavirus, Politics | pandemic, realdonaldtrump, biden, americans, states, joebiden, house, white, donald, testing, trumps, gop, america, administration, vote |
| 31 | Impeachment | impeachment, senate, house, trial, gop, witnesses, trumps, bolton, ukraine, republicans, white, senators, barr, vote, parnas |
| 32 | Trump Train, Trump Support Accounts | kbusmc2, jcpexpress, nobodybutme17, swilleford2, davidf4444, realdonaldtrump, tee2019k, amateurmmo, theycallmedoc1, rebarbill, sam232343433, kimber82604467, pawleybaby1999, thegrayrider, unyielding5 |
| 33 | US 2020 Politics | realdonaldtrump, democrats, bernie, america, democrat, bloomberg, pelosi, biden, impeachment, obama, house, american, vote, potus, dems |
| 34 | India Politics, Covid | india, delhi, narendramodi, pm, ani, govt, indian, police, lockdown, modi, minister, bjp, ji, positive, corona |
| 35 | Covid News, Covid Trackers, Itals | wuhan, chinese, outbreak, confirmed, case, italy, bnodesk, death, reuters, deaths, infected, reports, hospital, patients, spread |
| 36 | Nigeria Politics, Covid | nigeria, lagos, buhari, –, nigerian, lockdown, god, nigerians, mbuhari, govt, governor, ncdcgov, africa, police, money |
| 37 | Lifestyle, Art | □, love, ❤️, happy, be, morning, got, beautiful, night, birthday, week, music, friends, art, little |
| 38 | Australian Climate Change, Wildfires | australia, morrison, australian, climate, auspol, scottmorrisonmp, change, fire, nsw, scott, minister, pm, fires, bushfires, bushfire |
| 39 | Conservative Canadian Politics | trudeau, canada, canadians, canadian, ezralevant, justintrudeau, justin, liberal, liberals, police, un, minister, climate, cbc, alberta |
| 40 | UK Covid, Politics | uk, nhs, johnson, boris, care, labour, eu, £, brexit, lockdown, bbc, borisjohnson, deaths, staff, workers |
| 41 | Pro-Trump, Anti-Trump Accounts, QAnon | cspanwj, ngirrard, nevertrump, realdonaldtrump, govmlg, freelion7, tgradous, wearethenewsnow, nm, missyjo79, durango96380362, epitwitter, newmexico, potus, jubilee7double |
| 42 | Fox News, HCQ | zevdr, realdonaldtrump, niro60487270, foxnewspolitics, biden, hotairblog, , sierraamv, pandemic, america, be, dr, joe, obama, democrats |
| 43 | Hong Kong Protests, China Covid | hong, kong, chinese, strike, medical, hongkong, workers, border, huawei, police, hk, outbreak, hospital, wuhan, communist |
| 44 | Cuba and Venezuela Politics, Yang Support | cuba, venezuela, ●, andrewyang, \|, diazcanelb, —, cuban, maduroen, ┐, telesurenglish, yang, maduro, yanggang, pandemic |
| 45 | Animal Rescue | ❤️, , dog, dogs, animals, maryjoe38642126, animal, pls, ⚠️, cat, shelter, rescue, cats, needs, keitholbermann |
| 46 | US General Politics | realdonaldtrump, obama, donald, americans, house, gop, trumps, pandemic, joebiden, white, biden, america, dr, be, states |
| 47 | Left-Leaning, US and NYC Covid Equipment Shortages | pandemic, crisis, realdonaldtrump, americans, medical, masks, ventilators, hospital, york, workers, bill, dr, cnn, cuomo, hospitals |
| 48 | Trump Support, Bolsonaro Support | realdonaldtrump, mrfungiq, loveon70, jeffmctn1, monster4341, brazil, bolsonaro, 1technobuddy, basedpoland, followed, wickeddog3, brazilian, jairbolsonaro, 6831bryan, bonedaddy76 |
| 49 | US QAnon | realdonaldtrump, obama, biden, flynn, democrats, joe, fbi, america, obamagate, american, house, bill, fauci, realjameswoods, dr |